\documentclass[pra,aps,twocolumn,superscriptaddress]{revtex4-1}
\usepackage{amsfonts,amsmath,bm}
\usepackage{xcolor}
\usepackage{hyperref}

\newcommand{\rr}{\bm{r}}
\newcommand{\kk}{\bm{k}}
\newcommand{\RR}{\bm{R}}
\newcommand{\kp}{\mbox{$\bm{k}\!\cdot\!\bm{p}$} }

\usepackage{graphicx}

\begin{document}

\title{Phonon-assisted carrier tunneling with
  hyperfine-induced spin flip in coupled quantum dot systems}
\author{Pawe{\l} Karwat}
\affiliation{School of Physics and CRANN Institute, Trinity College Dublin, Dublin 2,
  Ireland}
\affiliation{Department of Theoretical Physics, Wroc{\l}aw University of Science and
	Technology, 50-370 Wroc{\l}aw, Poland} 
\author{Krzysztof Gawarecki}
\author{Pawe{\l} Machnikowski}
\affiliation{Department of Theoretical Physics, Wroc{\l}aw University of Science and
  Technology, 50-370 Wroc{\l}aw, Poland}

\begin{abstract}
We calculate the rates of phonon-assisted hyperfine spin flips during electron and hole
tunneling between quantum dots in a self-assembled quantum dot molecule. We show that the hyperfine
process dominates over the 
spin-orbit-induced spin relaxation in magnetic fields up to a few Tesla for electrons, while
for holes this cross-over takes place at field magnitudes of a fraction of Tesla, upon the
assumption of a large $d$-shell admixture to the valence band state, resulting in a strong
transverse hyperfine coupling. The interplay of the two spin-flip mechanisms leads to a
minimum of the spin-flip probability, which is in principle experimentally measurable and
can be used as a test for the presence of substantial transverse hyperfine couplings in
the valence band.
\end{abstract}

\maketitle

\section{Introduction}
\label{sec:intro}

Hyperfine coupling between carrier spins and nuclear magnetic moments
in a crystal \cite{Glazov-book} is one of the key factors determining the properties of semiconductor
structures, like quantum dots (QDs), and their usability in new information processing
devices. Once considered the main source of dephasing 
\cite{Schliemann_JPC03,Coish_PSSB09,Cywinski_APPA11,Urbaszek2013,Chekhovich2011c}, it
can now be controlled with growing precision and used as a manageable degree of
freedom~\cite{Smirnov2020,Chekhovich2020,Waeber2019}. 

The hf coupling for electrons is dominated by the approximately isotropic contact
interaction. Its transverse components can in principle lead to spin relaxation
accompanied by a simultaneous change of one of the nuclear spins (a `flip-flop'
process). Due to a large mismatch between electronic and nuclear Zeeman energies, this
process is restricted to very low magnetic fields
\cite{Eble2009a,Dou2012,Fras2012a} or bright-dark exciton resonances \cite{Kurtze2012}. At
magnetic fields exceeding a fraction of Tesla the hf processes become ineffective
\cite{Braun_PRL05,Glasenapp2016}. Even though
the energy gap can be closed by emitting a phonon \cite{Beyer2012a}, it turns out that
processes relying on the  spin-orbit (SO) coupling dominate in this regime
\cite{heiss07,bulaev05a} due to their more favorable dependence on the magnetic field. 
A cross-over between the hf and SO regimes has been observed in gated GaAs QDs \cite{Camenzind2018}. 

Theoretical description of hf-induced spin flips was developed in the context of electrons
in gated QDs
\cite{erlingsson02,Erlingsson_PRB04}: One considers corrections to the carrier state with
a given spin, mediated by the hyperfine coupling with all the nuclei, treated as a
perturbation. In this way, the original state gets an admixture of inverted spin, which
allows transitions to a state with a nominally opposite spin via spin-conserving phonon
couplings. 
For a transition within the Zeeman doublet, the combination of the $\propto B^{-2}$ scaling
of the hf admixture (stemming from the carrier Zeeman energy, while the nuclear Zeeman
splitting is negligible), frequency dependence of the phonon spectral density, and van Vleck
cancellation leads to a $\propto B^{3}$ dependence on the magnetic field.

In the case of holes the physics of hf interactions is more complex and some questions
seem to remain open. Overall, the hole hf coupling is due to dipole interactions, which
renders it much weaker than
the contact interaction of electrons \cite{Fischer2008,Testelin2009a,Fallahi2010}.
Moreover, for a purely 
$p$-type valence band, the transverse components of the hole hf coupling can only result
from weak band mixing effects 
\cite{Fischer2008,Eble2009a,Testelin2009a}, which would strongly limit hole spin
relaxation. Indeed, a coherent population trapping experiment under transverse nuclear spin
polarization in a low-noise device \cite{Prechtel2016} has led
to the conclusion that the transverse hyperfine coupling is negligible.
On the other hand, selective measurements of the Overhauser field
for particular elements and isotopes in the crystal \cite{Chekhovich2013} yield results
that can be explained by a substantial admixture of atomic $d$-shell 
states to the valence band, in line with earlier theoretical calculations
\cite{Boguslawski1994}. According to theoretical models, such a $d$-shell admixture would
give rise to a substantial transverse contribution to the hole hyperfine coupling
\cite{Machnikowski2019}.  

In quantum dot molecules (QDMs), built of two coupled QDs, even in the $s$ shell, spin
relaxation can take place not only between states within one Zeeman doublet but can also
accompany charge relaxation (dissipative tunneling) between the QDs.
Understanding the role of hyperfine interactions in such a carrier tunneling process may
be important for possible spin injection schemes as well as for spin readout protocols
involving carrier transfer induced by gating pulses. On the other hand, hf flip-flops
combined with spatio-temporal dynamics of the carrier may be used to imprint a particular
state in the nuclear system. Finally, theoretically predicted characteristics of the spin
relaxation, when confronted with the experiment, may verify the assumptions of the model
and thus offer information on the nature of the hf coupling itself.

A reliable description
of processes involving tunneling in self-assembled QD systems requires reasonable knowledge of
wave functions, which is achievable with the \kp method in the envelope
function approximation \cite{Gawarecki2012}. The \kp method allows one also to describe
all kinds of phonon-assisted processes, including those involving SO-induced spin
flips \cite{Mielnik-Pyszczorski2018a}.  Recently, we have also combined the \kp model with
the hyperfine Hamiltonian and provided a description of hyperfine couplings based on
multi-band wave functions \cite{Machnikowski2019}.

In this paper we calculate the rates of phonon-assisted hyperfine spin flip-flops during
electron and hole relaxation between the two branches of the $s$ shell in a QDM
(corresponding to states localized in two different QDs if the system is away from the
resonance). We compare the result with the SO-induced phonon-assisted spin-flip
tunneling and show that the hyperfine process dominates for fields below  a few
Tesla for electrons (depending on the axial electric field), while for holes it becomes
important only for fields roughly below 0.1~T. The interplay of hf and SO mechanisms of
spin relaxation leads to a non-monotonic dependence of the total spin flip probability,
which may be used as a test for considerable transverse hyperfine couplings.

The paper is organized as follows. In Sec.~\ref{sec:model} we desccribe the model of the
system under study. Sec.~\ref{sec:theory} presents the theory for the phonon-assisted
tunneling of a carrier with a
simultaneous spin flip-flop between the carrier and a nuclear spin. In
Sec.~\ref{sec:results} we present the results. Sec.

\section{Model}
\label{sec:model}

\begin{figure}[tb]
	\begin{center}
		\includegraphics[width=90mm]{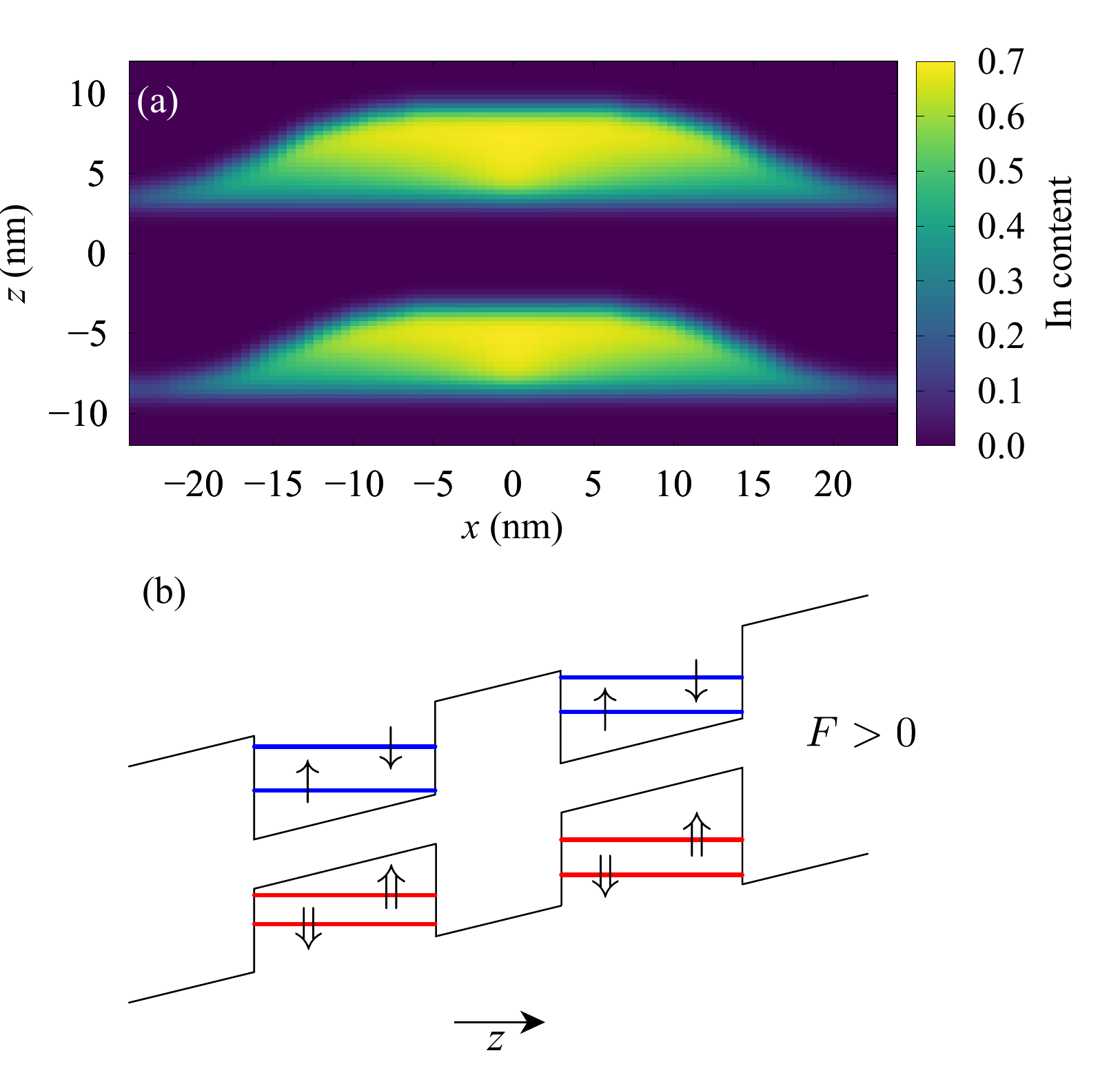}
	\end{center}
	\caption{\label{fig:comp}\textcolor{gray}({Color online) Material distribution in the system (a) and scheme of the lowest energy levels (b). } }
\end{figure}
We consider two coupled, vertically stacked self-assembled InGaAs quantum dots. The dots
are placed on the wetting layers of $a_G = 0.565325$~nm thickness and
$\mathrm{In_{0.4}\mathrm{Ga}_{0.6}As}$ composition. The shape of each dot is defined by
the restriction of its upper limit by the surface \cite{Ardelt2016a}
$$
S_{n}(\rr) = z^{\mathrm{(WL)}}_{n} + h_{n} \exp \left ( - \frac{(x^2 + y^2)^2}{r^4_{n}}  \right ),
$$
where $n=\{1,2 \}$ refers to the bottom and the top dot respectively,
$z^{\mathrm{(WL)}}_{n}$ denotes the top of the wetting layer, $h_{n}$ is the dot height,
and $r_{n}$ defines the lateral extension. We take $h_{1} = 4.8$~nm, $h_{2} = 5.4$~nm,
$r_{1}=15.8$~nm, and $r_{2}=17$~nm. Both dots have a trumpet-shape composition
\cite{jovanov12}, where the position-dependent In content is given by 
\begin{align*}
 C(\rr) = & ~C_{\mathrm{b}} + (C_{\mathrm{t}}-C_{\mathrm{b}}) \exp{ \left( \frac{
           -\sqrt{x^{2}+y^{2}} \exp{(-z/z_{\mathrm{c}})} }{r_{\mathrm{c}}}  \right )}, 
\end{align*}
where $C_{\mathrm{t}} = 0.7$, $C_{\mathrm{b}} = 0.4$ are related to the composition in the
top/bottom region of a given QD, $z_{\mathrm{c}} = 1.1$~nm and $r_{\mathrm{c}} = 2.3$~nm. To
simulate material intermixing, the dots are processed by a Gaussian blur with the standard
deviation of $0.6$~nm. The computations were carried out on $(200\times200\times200) a_G$ and $(100\times100\times70) a_G$ computational boxes for calculation of strain and electron states, respectively.

In order to quantitatively account for the hyperfine interactions in an inhomogeneous
system like a QD, one needs to model the properties of wave functions on the mesoscopic
scale of several or tens of nanometers and, simultaneously, describe the properties of
Bloch functions near the nuclei, on a sub-nanometer scale. No single computational method
is currently able to bridge these two scales, hence we use the hybrid approach developed
in Ref.~\cite{Machnikowski2019}, in which the mesoscopic scale is covered by an 8-band \kp
theory in the virtual crystal approximation, while the atomic-scale properties are
approximately accounted for by a simple 
model of effective rescaled hydrogen-like orbitals with averaging over specific alloying
and isotope configurations. 

Thus, on the mesoscopic scale, the static strain related to the lattice mismatch is
accounted for within the continuous 
elasticity approach \cite{pryor98b}. The strain-induced piezoelectric potential is
calculated including the polarization up to the second order in strain tensor
elements \cite{bester06b} with parameters from Ref.~\cite{caro15}. The wave functions
for the four lowest electron and hole states (corresponding to the two QDs and two spin
orientations) are obtained using the eight band \kp theory in the envelope function
approximation. The Hamiltonian, the material parameters and implementation is
given in much detail in Ref.~\cite{Gawarecki2018}. We used also improvements of the model described in Ref.~\cite{Krzykowski2020}.

The hyperfine Hamiltonian for the interaction of a carrier with nuclear spins is
\begin{equation}\label{Hhf}
H_{\mathrm{hf}}=3E_{\mathrm{hf}}
\sum_{\alpha} \zeta_{\alpha}
\bm{A}(\rr-\RR_{\alpha})\cdot\bm{I}_{\alpha}/\hbar,
\end{equation}
where $\alpha$ labels the ions, $\RR_{\alpha}$ are their positions,
\begin{displaymath}
E_{\mathrm{hf}} =
\frac{2\mu_{0}}{3\pi}\mu_{\mathrm{B}}\mu_{\mathrm{N}}
a_{\mathrm{B}}^{-3} = 0.5253\,\mu\mathrm{eV},
\end{displaymath}
$\mu_{\mathrm{B}}$ and $\mu_{\mathrm{N}}$ are Bohr and nuclear magnetons, respectively,
$a_{\mathrm{B}}$ is the Bohr radius, $\mu_{0}$ is the vacuum permeability,
$\bm{I}_{\alpha}$ is the nuclear spin, $\zeta_{\alpha}$ defines the nuclear
magnetic moment for a given nucleus via
$\bm{\mu}_{\alpha}=\zeta_{\alpha}\mu_{\mathrm{N}}\bm{I}_{\alpha}$, and
\begin{equation}\label{A-all}
\bm{A}(\rr) = \frac{a_{\mathrm{B}}^{3}}{4\hbar}\left[
\frac{8\pi}{3}\delta(\rr)\bm{S} + \frac{\bm{L}}{r^{3}}
+\frac{3(\hat{\rr}\cdot\bm{S})\hat{\rr}-\bm{S}}{r^{3}}  \right].
\end{equation}
We consider the two most common In isotopes, two Ga isotopes and one As isotope with the
angular momenta $j=$9/2, 3/2, and 3/2, respectively.

\begin{table}
\caption{\label{params}Nuclear \cite{Schliemann_JPC03} and atomic
parameters.}
\begin{center}
\begin{tabular}{l|ccccc}
 & $^{69}$Ga & $^{71}$Ga & $^{113}$In & $^{115}$In & $^{75}$As \\
\hline
$I$ & $3/2$  & $3/2$ & $9/2$ & $9/2$ & $3/2$ \\
$\zeta$ & 1.344 & 1.708 &  1.227 & 1.230 & 0.959\\
%$I\zeta_{\alpha}$ & 2.016 & 2.562 & 5.523 & 5.534 & 1.439 \\
$r$ & 0.604 & 0.396 & 0.0428 & 0.9572 & 1 \\
\hline
$\xi_{s}$ & \multicolumn{2}{c}{3.9} & \multicolumn{2}{c}{3.9}  & 4.4 \\
$\xi_{p}$ & \multicolumn{2}{c}{3.3} & \multicolumn{2}{c}{3.3} & 3.7 \\
$\xi_{d}$ & \multicolumn{2}{c}{10.5} &  \multicolumn{2}{c}{8.9} & 11.9 \\
\hline
$M_{p}$ & \multicolumn{2}{c}{0.050} & \multicolumn{2}{c}{0.050} & 0.050 \\
$M_{d}$ & \multicolumn{2}{c}{0.33} & \multicolumn{2}{c}{0.20} & 0.33 \\
$M_{sd}$ & \multicolumn{2}{c}{0.048} &  \multicolumn{2}{c}{0.034} & 0.049 \\
\hline
$|\alpha_{d}|^{2}$ & \multicolumn{2}{c}{0.20} &  \multicolumn{2}{c}{0.50} & 0.05 \\
\hline
$|a_{{\mathrm{C/A}}}^{\mathrm{cb}}|^{2}$ & \multicolumn{4}{c}{0.50} & 0.50 \\
$|a_{{\mathrm{C/A}}}^{\mathrm{vb}}|^{2}$ & \multicolumn{4}{c}{0.35} & 0.65 \\
\end{tabular}
\end{center}
\end{table}

 The Bloch function is modeled as a
sum of atomic orbitals corresponding to the outermost shells, centered around the two
nuclei in the primitive cell. The atomic 
orbitals are taken as hydrogen-like wave functions 
$\psi_{nlm}(\rr) = \sqrt{n\xi_{nl}}\psi^{(\mathrm{H})}_{nlm}(n\xi_{nl}\rr)$, where
$\psi^{(\mathrm{H})}_{nlm}(\rr)$ are the wave functions of the hydrogen atom and $\xi_{nl}$
is a scaling parameter. Since only one shell of a given symmetry is relevant for the
topmost valence and lowest conduction bands, the principal quantum number $n$ will be
omitted. The scaling parameters $\xi_{s}$ for the $s$-shell states are obtained from the
experimentally determined values of the conduction band wave functions at the nucleus
\cite{Chekhovich2017a,Gueron1964}. The distribution of the wave functions between the
anion and cation is chosen to be consistent with the known experimental
\cite{Chekhovich2017a,Gueron1964} and computational \cite{Boguslawski1994} data. In view
of the lack of precise data, we choose $\xi_{p}=0.85\xi_{s}$ for each atom, following the
general relation between the scaling parameters of Slater orbitals
\cite{Clementi1963,Clementi1967,Benchamekh2015} (neither the Slater orbitals themselves
nor their scaling commonly used in tight-banding computations can be used directly, as
they are optimized against chemical bonding and band structures and fail to reproduce the
correct behavior near the nucleus). The $d$-shell scaling parameter $\xi_{d}$ is estimated
from the measured values of the hf coupling for holes \cite{Chekhovich2013} using the
theoretically computed weights of $d$-shell admixtures in GaAs \cite{Boguslawski1994}.
The resulting values of the parameters describing the hyperfine coupling are listed in 
Tab.~\ref{params}, where we also give the quantities $M_{ll'} = |\psi_{s}(0)|^{-2}
\langle l | 1/r^{3} | l' \rangle$ that characterize the geometry of the wave functions for
the dipole hyperfine interaction. The details of the model are described in 
 Ref.~\cite{Machnikowski2019}. 

In view of the very small value of the nuclear Zeeman splitting, the 
probability for any nuclear configuration at thermal equilibrium is essentially the same
at typical temperatures of experiments involving self-assembled QDs (a few Kelvin and above).
Since the carrier Zeeman splitting is much larger than the nuclear one, the nuclear Zeeman
energies can be neglected when considering the energy change in a carrier-nucleus spin
flip-flop. 

Coupling to phonons is described in the usual way.
The phonon subsystem and the
carrier-phonon interaction are described by the general Hamiltonian
\begin{displaymath}
H_{ph} = \sum_{\kk,\lambda}\hbar\omega_{\kk,\lambda}
b_{\kk,\lambda}^{\dag}b_{\kk,\lambda}
+ \sum_{\kk,\lambda} \left\{\Phi(\rr),e^{i\kk\cdot\rr} \right\}\left(
b_{\kk,\lambda}+b_{-\kk,\lambda}^{\dag} \right),
\end{displaymath}
where $\kk$ and $\lambda$ denote the wave vector and polarization of a phonon mode,
respectively, $b_{\kk,\lambda}$, $b_{-\kk,\lambda}^{\dag}$ are the corresponding
annihilation and creation operators, and $\Phi(\rr)$ is an $8\times 8$ matrix of
operators in the coordinate representation, corresponding to the 8-band structure of the
\kp theory and accounting for deformation-potential and piezoelectric couplings. The
detailed description of the carrier-phonon Hamiltonian is given in 
Ref.~\cite{Gawarecki2019a}. 

\section{Phonon-assisted spin-flip transitions}
\label{sec:theory}

In this section we present the theory for the phonon-assisted tunneling of carriers
between the ground state manifolds of the two QDs with a simultaneous spin flip-flop
between the carrier and a nuclear spin. The 8-band \kp theory treats the electron and hole
states on equal footing (the latter upon a standard transition from the electron picture of
the completely filled valence band to the hole picture) and we present our theory for
single-particle states in a general form, without specifying the kind of the carrier. In
the following, the term ``spin'' is used to identify one of the two sub-bands of the
conduction or 
heavy-hole valence band.

In order to find the rate for a spin flip-flop process we first calculate the hyperfine
flip-flop correction to the system 
state. We denote the carrier state in the $n$th QD ($n=1,2$) with the nominal spin
orientation $\sigma$ (resulting from the \kp diagonalization) by $|n\sigma\rangle$ and its
energy by $E_{\sigma}^{(n)}$. The state of the nuclei is labeled by
$|\ldots m_{\alpha}\ldots \rangle$, where $m_{\alpha}$ is the quantum number for the $z$ projection
of the respective nuclear spin. The states unperturbed by the hyperfine coupling are of
a product form $|n\sigma;\ldots m_{\alpha}\ldots\rangle =
 |n\sigma \rangle \otimes |\ldots m_{\alpha}\ldots \rangle$.
In the lowest order of perturbation theory with respect to the hyperfine interaction the
eigenstates of the system are then
\begin{align} \label{pert}
|\Psi_{n\sigma;\ldots m_{\alpha}\ldots}\rangle & = |n\sigma;\ldots m_{\alpha}\ldots\rangle \\
& \quad+\sum_{\alpha}c_{\alpha+}^{(n\sigma)} |n\overline{\sigma}; \ldots
  m_{\alpha}+1\ldots\rangle \nonumber\\
& \quad +\sum_{\alpha}c_{\alpha-}^{(n\sigma)} |n\overline{\sigma}; \ldots
  m_{\alpha}-1\ldots\rangle, \nonumber
\end{align}
where $\overline{\sigma}$ denotes inverted spin and the nuclear configuration on the
right-hand side is the same as the one on the left-hand side except for the one explicitly
given modified spin.
The coefficients of the perturbative correction are
\begin{align*}
c_{\alpha\pm}^{(n\sigma)}  &= \frac{
\langle n\overline{\sigma}; \ldots m_{\alpha}\pm 1\ldots | H_{\mathrm{hf}}
|n\sigma; \ldots m_{\alpha}\ldots\rangle
}{\hbar\omega_{\sigma \overline{\sigma}}^{(nn)}} \\
& = \frac{3E_{\mathrm{hf}} \zeta_{\alpha}}{2\hbar \omega_{\sigma \overline{\sigma}}^{(nn)}}
\sqrt{j(j+1)-m_{\alpha}(m_{\alpha}\pm 1)}
\left\langle n\overline{\sigma} \left| A_{\mp} \right| n\sigma \right\rangle,
\end{align*}
where $A_{\pm} = A_{x}\pm iA_{y}$ and 
$\omega_{\sigma\sigma'}^{(nn')}  = (E_{\sigma}^{(n)}-E_{\sigma'}^{(n')})/\hbar$.

From the Fermi golden rule, the probability of a phonon-assisted transition from the
state with spin $\sigma$ in QD1 to the state with spin $\sigma'$ in QD2 with a change of
the nuclear configuration from $\{m_{\alpha}\}$ to $\{m'_{\alpha}\}$ is
\begin{align*}
\lefteqn{\Gamma_{\sigma\to\sigma'} ^{\{m_{\alpha}\} \to \{m'_{\alpha}\}}=} \\
&\quad \frac{2\pi}{\hbar}
\left| n_{\mathrm{B}}(\omega_{\sigma\sigma'}^{(12)})+1 \right|\sum_{\kk,\lambda}
\delta(\hbar\omega_{\kk,\lambda}-|\omega_{\sigma\sigma'}^{(12)}|) \\
&\quad\times
\left| \left\langle \Psi_{2\sigma';\ldots m'_{\alpha}\ldots} |
    \left\{\Phi(\rr),e^{i\kk\cdot\rr} \right\}
 | \Psi_{1\sigma;\ldots  m_{\alpha}\ldots} \right\rangle \right|^{2}.
\end{align*}
Substituting the perturbation expansion from Eq.~\eqref{pert}, taking into account the
obvious fact that the phonon interaction conserves nuclear  spins, and denoting
\begin{equation}\label{Fk}
F_{\sigma'\sigma}(\kk) = \left\langle 2\sigma' |
    \left\{\Phi(\rr),e^{i\kk\cdot\rr} \right\} | 1\sigma \right\rangle
\end{equation}
one finds
\begin{align}\label{amplitude}
\lefteqn{ \left\langle \Psi_{2\sigma';\ldots m'_{\alpha}\ldots} |
    \left\{\Phi(\rr),e^{i\kk\cdot\rr} \right\}
 | \Psi_{1\sigma;\ldots  m_{\alpha}\ldots} \right\rangle =} \\
&\quad F_{\sigma'\sigma}(\kk) \nonumber\\
&\quad + F_{\sigma'\overline{\sigma}}(\kk) \sum_{\alpha,\pm} c_{\alpha\pm}^{(1\sigma)}
\langle\ldots m_{\alpha}'\ldots | \ldots m_{\alpha}\pm 1 \ldots \rangle \nonumber\\
&\quad + F_{\overline{\sigma'}\sigma}(\kk) \sum_{\alpha,\pm} c_{\alpha\pm}^{(2\sigma')*}
\langle\ldots m_{\alpha}'\pm 1\ldots | \ldots m_{\alpha}\ldots \rangle \nonumber\\
& \quad + \ldots.
\end{align}
Since the multi-band carrier wave functions are dominated by one leading component
(determining the nominal ``spin'' of the state), the couplings $F_{\sigma'\sigma}(\kk)$
are large for $\sigma=\sigma'$ and much smaller otherwise, when they stem from the band
mixing involving SO couplings. The hyperfine 
admixture amplitudes $c_{\alpha\pm}^{(n\sigma)}$ are small, as well.
Therefore, in Eq.~\eqref{amplitude} we kept only the contributions in the leading order in
the SO or hf couplings, neglecting those relying on both these weak
couplings simultaneously. Furthermore, for a nominally spin-conserving
process ($\sigma=\sigma'$),  the transition amplitude is by far dominated by the first
contribution $F_{\sigma\sigma}(\kk)$, which determines the spin-conserving phonon-assisted
tunneling rate
\begin{equation}\label{cons}
\Gamma_{\sigma\to\sigma} =
2\pi R_{\sigma\sigma\sigma\sigma}(|\omega_{\sigma\sigma}^{(12)}|),
\end{equation}
where we define the spectral densities for the phonon bath as
\begin{align}\label{R}
R_{\sigma_{1}\sigma_{2}\sigma_{3}\sigma_{4}}(\omega) &= \frac{1}{\hbar^{2}}
\left| n_{\mathrm{B}}(\omega)+1 \right| \\
&\quad \times\sum_{\kk,\lambda}
F_{\sigma_{1}\sigma_{2}}(\kk) F_{\sigma_{4}\sigma_{3}}^{*}(\kk)
\delta(\omega_{\kk,\lambda}-|\omega|). \nonumber
\end{align}

For a spin-flip process, there are two mechanisms that may, in principle, yield
comparable contributions: the SO channel entering via the first term and the
hyperfine channel accounted for by the two other terms on the right-hand side of
Eq.~\eqref{amplitude}. The total spin-flip transition rate is then a sum of the SO rate,
\begin{equation}\label{s0}
\Gamma_{\sigma\to\overline{\sigma}}^{(\mathrm{so})} =
2\pi R_{\sigma\overline{\sigma}\overline{\sigma}\sigma}(\omega_{\sigma\overline{\sigma}}^{(12)}),
\end{equation}
and the rates for hyperfine transitions, summed up over final
configurations of the nuclear bath, differing by one nuclear spin-flip from the initial one,
\begin{equation}\label{hf}
\Gamma_{\sigma\to\overline{\sigma}}^{(\mathrm{hf})} (\ldots m_{\alpha}\ldots) =
 \sum_{\alpha} \Gamma_{\sigma\to\overline{\sigma}}^{(\mathrm{hf}),\alpha} (\ldots m_{\alpha}\ldots),
\end{equation}
where we explicitly noted the dependence on the initial configuration of the nuclear bath
and
\begin{equation}\label{hf-alfa}
\Gamma_{\sigma\to\overline{\sigma}}^{(\mathrm{hf}),\alpha} (\ldots m_{\alpha}\ldots) =
2\pi \sum_{a=\sigma,\overline{\sigma}}\sum_{b=\sigma,\overline{\sigma}}
Q_{ab}^{(\alpha)}
R_{aabb} (\omega_{\sigma\overline{\sigma}}^{(12)}),
\end{equation}
with
\begin{widetext}
\begin{align*}
Q_{\sigma\sigma}^{(\alpha)} & =
\left( \frac{3E_{\mathrm{hf}} \zeta_{\alpha}}{2\hbar \omega_{\overline{\sigma}\sigma}^{(22)}} \right)^{2}
\left\{
\left[ j(j+1)-m_{\alpha}(m_{\alpha}-1) \right] \left|
\left\langle 2\sigma \left| A_{-} \right| 2\overline{\sigma} \right\rangle\right|^{2}
+\left[ j(j+1)-m_{\alpha}(m_{\alpha}+1) \right] \left|
\left\langle 2\sigma \left| A_{+} \right| 2\overline{\sigma} \right\rangle\right|^{2}
\right\}, \\
Q_{\overline{\sigma}\overline{\sigma}}^{(\alpha)} & = \left(
\frac{3E_{\mathrm{hf}} \zeta_{\alpha}}{2\hbar \omega_{\sigma \overline{\sigma}}^{(11)}} \right)^{2}
\left\{
\left[ j(j+1)-m_{\alpha}(m_{\alpha}+1) \right] \left|
\left\langle 1\overline{\sigma} \left| A_{-} \right| 1\sigma \right\rangle\right|^{2}
+\left[ j(j+1)-m_{\alpha}(m_{\alpha}-1) \right] \left|
\left\langle 1\overline{\sigma} \left| A_{+} \right| 1\sigma \right\rangle\right|^{2}
\right\}, \\
Q_{\overline{\sigma}\sigma}^{(\alpha)} &= Q_{\sigma\overline{\sigma}}^{(\alpha)*} =
\frac{(3E_{\mathrm{hf}} \zeta_{\alpha})^{2}}{4\hbar^{2}
\omega_{\sigma \overline{\sigma}}^{(11)}\omega_{\overline{\sigma}\sigma}^{(22)}}
\left\{
\left[ j(j+1)-m_{\alpha}(m_{\alpha}+1) \right]
\left\langle 1\overline{\sigma} \left| A_{-} \right| 1\sigma \right\rangle
\left\langle 2\sigma \left| A_{+} \right| 2\overline{\sigma} \right\rangle \right.\\
&\quad \left. +\left[ j(j+1)-m_{\alpha}(m_{\alpha}-1) \right]
\left\langle 1\overline{\sigma} \left| A_{+} \right| 1\sigma \right\rangle
\left\langle 2\sigma \left| A_{-} \right| 2\overline{\sigma} \right\rangle
\right\}.
\end{align*}
\end{widetext}
Note that
$R_{\sigma\sigma\overline{\sigma}\overline{\sigma}} (\omega)
= R_{\overline{\sigma}\overline{\sigma}\sigma\sigma}^{*}(\omega)$ so
$\Gamma_{\sigma\to\overline{\sigma}}^{(\mathrm{hf}),\alpha}$ is real.

Since the shape of the wave function for a given spatial state in a given QD very weakly
depends on the spin orientation, all the spectral densities in Eq.~\eqref{hf-alfa} are
almost identical upon an appropriate choice of the arbitrary phases. This allows one to
write
\begin{widetext}
\begin{align}\label{hf1}
\Gamma_{\sigma\to\overline{\sigma}}^{(\mathrm{hf}),\alpha} (\ldots
  m_{\alpha}\ldots) & =
2\pi R (\omega_{\sigma\overline{\sigma}}^{(12)})
\sum_{a=\sigma,\overline{\sigma}}\sum_{b=\sigma,\overline{\sigma}} Q_{ab}^{(\alpha)}
  \nonumber\\
& = 2\pi R (\omega_{\sigma\overline{\sigma}}^{(12)})
\frac{(3E_{\mathrm{hf}} \zeta_{\alpha})^{2}}{4}
\left\{
\left[ j(j+1)-m_{\alpha}(m_{\alpha}+1) \right]
\left|
\left\langle 2\sigma \left| A_{+} \right| 2\overline{\sigma} \right\rangle
- \left\langle 1\sigma \left| A_{+} \right| 1\overline{\sigma} \right\rangle
\right|^{2} \right. \nonumber\\
&\quad \left.
+\left[ j(j+1)-m_{\alpha}(m_{\alpha}-1) \right]
\left|
\left\langle 2\sigma \left| A_{-} \right| 2\overline{\sigma} \right\rangle
- \left\langle 1\sigma \left| A_{-} \right| 1\overline{\sigma} \right\rangle
\right|^{2} \right\}.
\end{align}
\end{widetext}

For a transition between the Zeeman states in a single QD the distinction between ``1''
and ``2'' disappears, the two matrix elements of
$A_{\pm}$ become identical, and the transition rate is suppressed by destructive
interference. In contrast, in the QDM system, as long as the states are spatially
separated (away from the level-crossing resonance at which the ground states in the two
QDs are aligned) each ion is  
effectively coupled to at most one carrier state (the one localized in the same QD as
the ion) and only one of the two interfering amplitudes can be large.

For unpolarized nuclei the physically meaningful rate is obtained by averaging
Eq.~\eqref{hf} over all the initial configurations of the nuclear spins and summing up
over all nuclear spin flips,
\begin{equation}\label{hf-aver}
\overline{\Gamma}_{\sigma\to\overline{\sigma}}^{(\mathrm{hf})} =
2\pi \sum_{a=\sigma,\overline{\sigma}}\sum_{b=\sigma,\overline{\sigma}}
\overline{Q}_{ab}
R_{aabb} (\omega_{\sigma\overline{\sigma}}^{(12)}).
\end{equation}
 Since
$\langle  j(j+1)-m_{\alpha}(m_{\alpha}\pm 1) \rangle=2j(j+1)/3$ one finds
\begin{subequations}
\label{Q-av}
\begin{align}\label{Q-av-a}
\overline{Q}_{\sigma\sigma} & = \sum_{\alpha}\left(
\frac{3E_{\mathrm{hf}} \zeta_{\alpha}}{2\hbar \omega_{\overline{\sigma}\sigma}^{(22)}} \right)^{2}
\frac{2j(j+1)}{3} \\
&\quad\times \left(
\left|\left\langle 2\sigma \left| A_{-} \right| 2\overline{\sigma} \right\rangle\right|^{2}
+ \left| \left\langle 2\sigma \left| A_{+} \right| 2\overline{\sigma}
                                  \right\rangle\right|^{2}
\right), \nonumber\\
\overline{Q}_{\overline{\sigma}\overline{\sigma}} & = \sum_{\alpha}
\left( \frac{3E_{\mathrm{hf}} \zeta_{\alpha}}{2\hbar \omega_{\sigma
                                                               \overline{\sigma}}^{(11)}}
                                                               \right)^{2}
\frac{2j(j+1)}{3} \\
& \quad\times \left(
\left| \left\langle 1\overline{\sigma} \left| A_{-} \right| 1\sigma \right\rangle\right|^{2}
+ \left| \left\langle 1\overline{\sigma} \left| A_{+} \right| 1\sigma \right\rangle\right|^{2}
\right), \nonumber\\
\overline{Q}_{\overline{\sigma}\sigma} =
  \overline{Q}_{\sigma\overline{\sigma}}^{*} &=
\sum_{\alpha}
\frac{(3E_{\mathrm{hf}} \zeta_{\alpha})^{2}}{4\hbar^{2}
\omega_{\sigma \overline{\sigma}}^{(11)}\omega_{\overline{\sigma}\sigma}^{(22)}}
\frac{2j(j+1)}{3} \\
\label{Q-av-c}
&\quad\times
\left(
\left\langle 1\overline{\sigma} \left| A_{-} \right| 1\sigma \right\rangle
\left\langle 2\sigma \left| A_{+} \right| 2\overline{\sigma} \right\rangle \right. \nonumber\\
&\quad \left. +
\left\langle 1\overline{\sigma} \left| A_{+} \right| 1\sigma \right\rangle
\left\langle 2\sigma \left| A_{-} \right| 2\overline{\sigma} \right\rangle
\right).\nonumber
\end{align}
\end{subequations}

\section{Results}
\label{sec:results}

In this section we analyze and compare the electron and hole phonon-assisted tunneling
rates with hyperfine-
and SO-induced spin flips, as well as the spin-conserving tunneling rates. The
rates are calculated using Eq.~\eqref{hf-aver}, corresponding to the thermal equilibrium
of the nuclear bath at temperatures much higher than the nuclear Zeeman energies.
The matrix elements in Eqs.~(\ref{Q-av}) are evaluated using the
\kp formalism for hyperfine interactions developed in our previous
paper \cite{Machnikowski2019}, while the 
spectral densities are computed directly from the \kp wave functions according to the
definitions in Eq.~\eqref{Fk} and Eq.~\eqref{R}.

\begin{figure}[tb]
	\begin{center}
		\includegraphics[width=80mm]{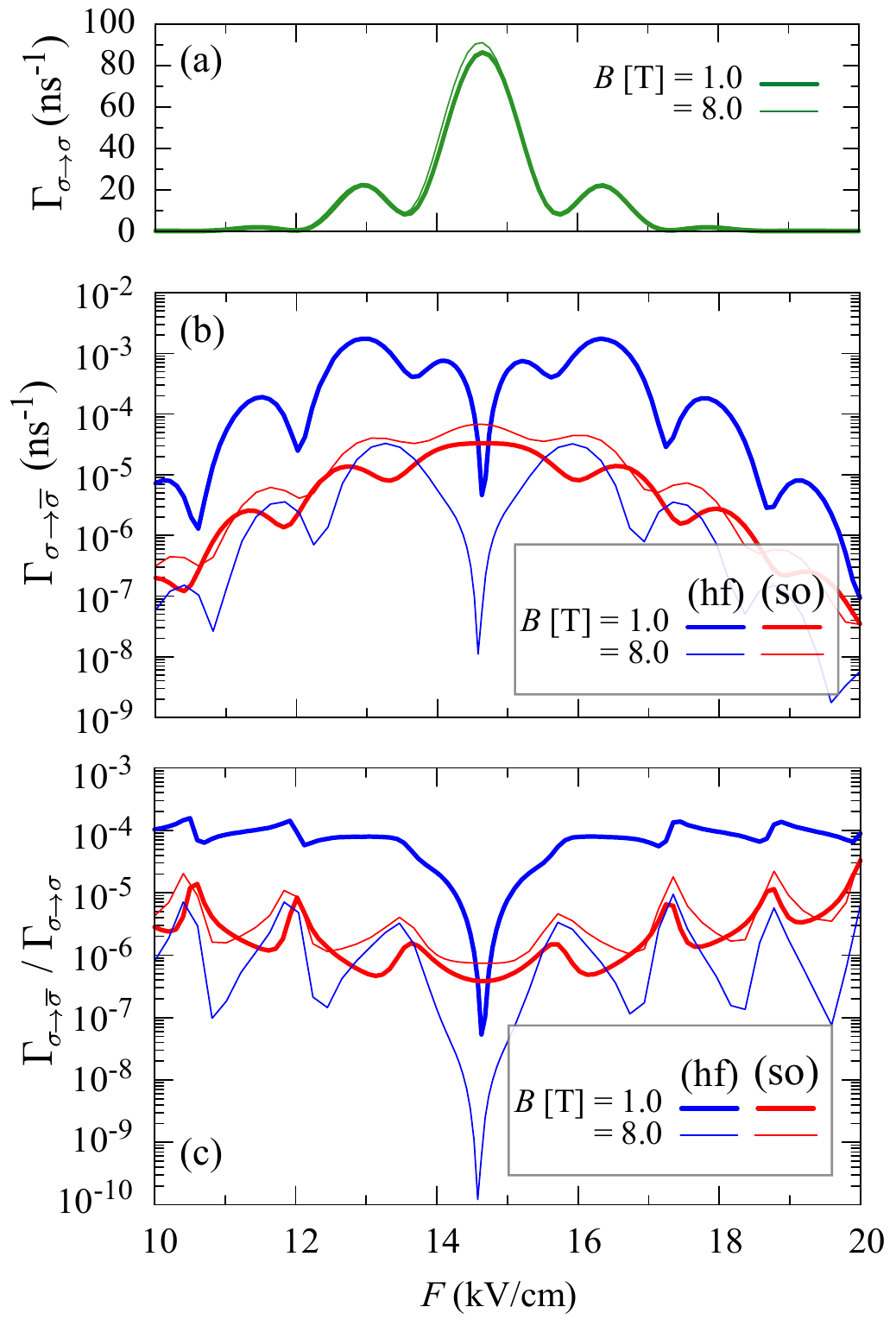}
	\end{center}
	\caption{\label{fig:g_combo_e}({Color online) (a) Spin-conserving phonon-assisted
            tunneling rate of an electron at two values of the magnetic field as a
            function of the axial electric field. (b) Spin-flip
            transition rates due the hf (blue lines) and 
            SO (red lines) interaction. (c) Ratio of the spin-flip to the
            spin-conserving phonon-assisted tunneling rate for both transition channels.}
        } 
\end{figure}

Fig.~\ref{fig:g_combo_e}(a) shows the spin-conserving relaxation rates between the two
lowest electron states as a function of the electric field $F$ applied in the growth
direction ($z$) that relatively shifts
the ground-state manifolds of the two QDs. The results are shown at two values of the
magnetic field oriented along the growth direction (Faraday configuration). The central
maximum corresponds to the tunneling 
resonance, when the two levels are aligned and form an anti-crossing. The oscillations are
due to the interplay between the QD separation and the wave length of the emitted phonon
as the energy splitting between the two levels is changed  \cite{gawarecki10}. The
difference between the 
relaxation rates at the two values of magnetic field is marginal. 

In Fig.~\ref{fig:g_combo_e}(b) we show the spin-flip tunneling rates for the electron in the
presence of electric and magnetic fields as previously, comparing the rates calculated
according to the theory presented above with those resulting from the SO couplings
\cite{Gaweczyk2019}. 
At weak magnetic fields (here we choose $B=1$~T) the hyperfine channel dominates over the
spin-orbit one. On the other hand,
for moderate and strong values of magnetic field (here $B=8.0$~T), the spin-flip transitions
caused by the SO couplings are stronger. This is due to the fact, that the rate
related to the hf interaction decreases like $\propto 1/B^{2}$, while for the
SO mechanisms it increases slowly in the considered range of the magnetic fields.
In the case of the
hyperfine-induced transitions, we observe a minimum at the field value corresponding to
the resonance condition (crossing of the ground states of the two QDs). Under such
conditions both wave functions are delocalized over the two QDs and a destructive
interference of the two matrix elements of $A_{\pm}$ in Eq.~\eqref{hf1} takes place. 

The ratio of
spin-flip to spin-conserving transition rates is shown in
Fig.~\ref{fig:g_combo_e}(c). This value is equal to the probability of spin flip during
the incoherent phonon-assisted tunneling and therefore is a measure of spin preservation
in this process. 
Typical values are on 
the order of $10^{-4}$ at 1~T, dominated by the hf coupling and scaling as $1/B^{2}$ at
lower fields, while at higher fields the spin-flip process is dominated by the SO coupling
and its probability is reduced to $10^{-6}$.
For the hf-induced process, the pattern of
oscillations in the spin-flip rate
$\overline{\Gamma}_{\sigma\to\overline{\sigma}}^{(\mathrm{hf})}$ is 
similar to that in the spin-conserving rate $\Gamma_{\sigma\to\sigma}$ and the relative
ratio is relatively flat. In
contrast, for the SO process, $\Gamma_{\sigma\to\overline{\sigma}}^{(\mathrm{so})}$ has a
different pattern 
of oscillations compared to $\Gamma_{\sigma\to\sigma}$, hence the ratio
$\Gamma_{\sigma\to\overline{\sigma}}^{(\mathrm{so})} /
\Gamma_{\sigma\to\overline{\sigma}}$ as a function of the electric field exhibits
pronounced maxima and minima. 

\begin{figure}[tb]
	\begin{center}
		\includegraphics[width=80mm]{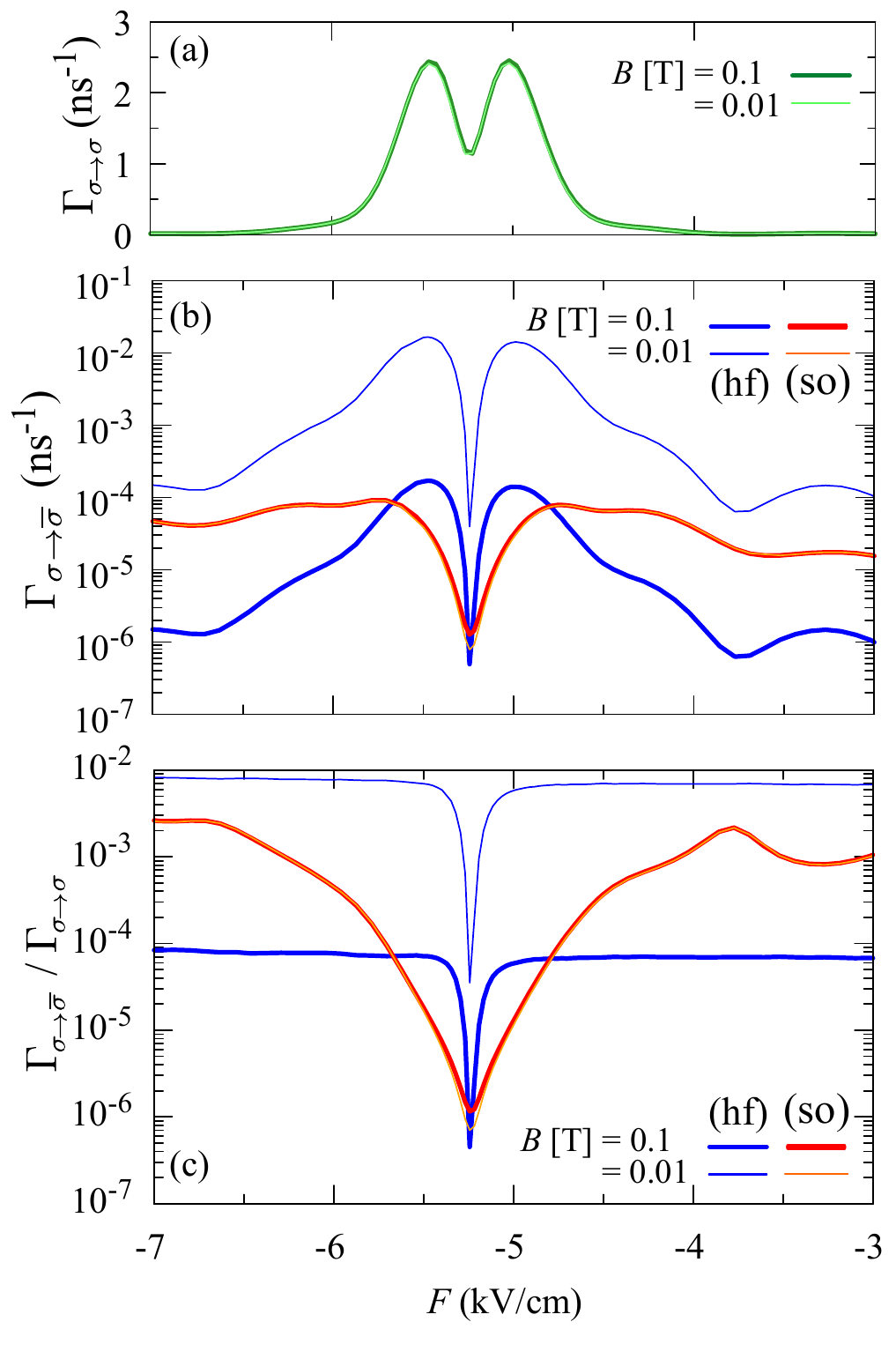}
	\end{center}
	\caption{\label{fig:g_combo_h}({Color online) (a) Spin-conserving phonon-assisted
            tunneling rate of a hole. (b) Spin-flip 
            transition rates related to the hf (blue lines) and SO (red lines)
            coupling. (c) Ratio of the spin-flip to 
            the spin-conserving phonon-assisted process for both transition
            channels.} }
\end{figure}

The results for a hole are presented in Fig.~\ref{fig:g_combo_h}. Again, we show the
electric field dependence of the spin-flip rate for 
the two spin relaxation mechanisms at two magnitudes of the magnetic field. For a hole,
the spin-conserving phonon-assisted relaxation, shown in Fig.~\ref{fig:g_combo_h}(a), is
slower than for an electron due to weaker deformation 
potential coupling. In addition, when the states are localized in different QDs, the
phonon-assisted tunneling is less efficient because of the higher hole mass, hence
stronger localization. 

For a hole, as can be seen in Fig.~\ref{fig:g_combo_h}(b), the two spin-flip channels
depend in different ways on the energy splitting controlled by the external electric
field. As a result, at~a~given magnetic field, one or the other process may dominate. In
the vicinity of the resonance, when the phonon-assisted relaxation or tunneling process is
effective, the cross-over between the two mechanisms occurs at magnetic field amplitudes
of a fraction of Tesla, much lower than in the case of an electron. This is due to the fact that 
the hyperfine-induced spin flip is much less probable than for an electron as a
consequence of the much lower hf coupling for holes,
while the SO-induced process is a few times more effective. Here the SO channel is nearly
magnetic-field-independent, while the rate for the hf channel scales as $1/B^{2}$.

The spin-flip probability, given by the ratio of the spin-conserving to spin-flip rates
[Fig.~\ref{fig:g_combo_h}(c)], is 
dominated by the hf-induced process at very low fields, on the order of 0.01~T and
below, with a remarkably high probability of the spin flip on the order of $1\%$ at
$B=0.01$~T except for the closest vicinity of the resonance. 
At higher fields the importance of this process is reduced, leading to a much lower
relative efficiency of the spin-flip process near the resonance, while far away from the
resonance the increasing SO contribution yield overall probabilities on the order of $10^{-3}$. 

\begin{figure}[h]
\begin{center}
  \includegraphics[width=87mm]{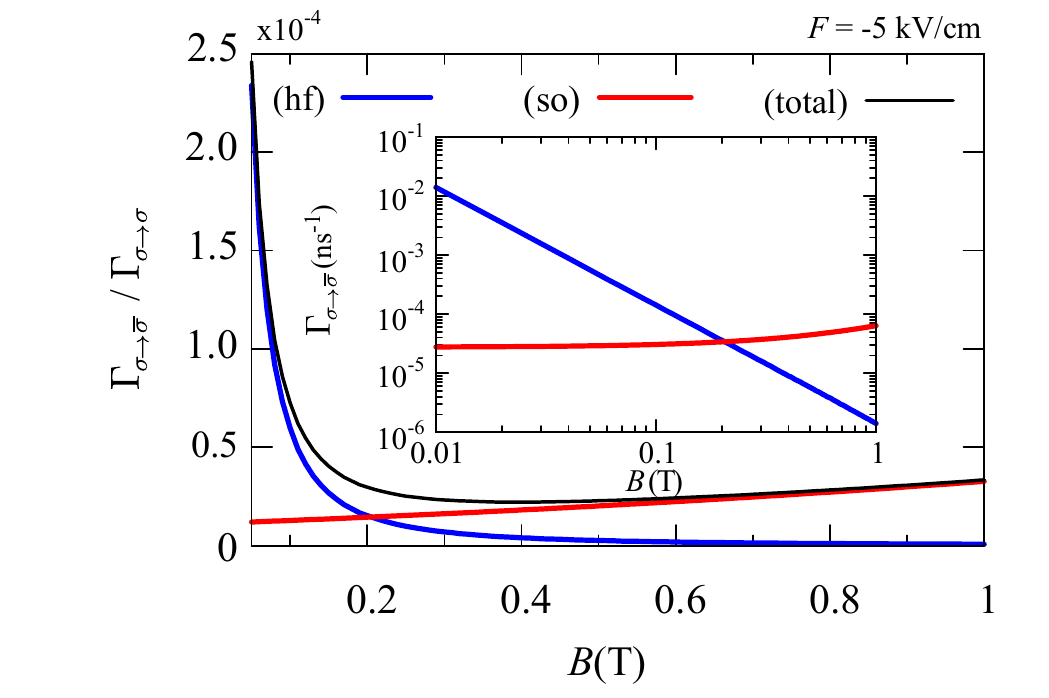}
\end{center}
\caption{\label{fig:F_h}(Color online) Spin-flip relaxation rates for a hole for the two
  spin-flip channels as a function of
  the magnetic field, for a fixed electric field $F=-5$~kV/cm. The relative spin-flip rates (the spin-flip probabilities) for the
  two channels along with the total relative rate. Inset: The absolute spin-flip
  relaxation rates for the two mechanisms.}
\end{figure}    

In Fig.~\ref{fig:F_h} we present the crossover between the SO and hf-dominated hole spin
flip as a function of the magnetic field at the electric field magnitude of $F=-5$~kV/cm,
near the maximum of the relaxation rate  (this time using a linear scale). At this
electric field value the cross--over takes 
place at $B\approx 0.2$~T (the exact value obviously depends on the electric field). While
the hf channel manifests the $1/B^{2}$ dependence visible already on the previous figure,
the SO contribution shows a very weak increase with the magnetic field (see the inset in
Fig. ~\ref{fig:F_h}), which was not 
noticeable earlier. As a result, the total relative spin-flip rate (spin-flip probability)
is a non-monotonic function of 
the magnetic field over an experimentally accessible range of field magnitudes.  
A similar dependence can also be observed for an electron but with a much less pronounced
minimum that is shifted to magnetic field magnitudes over 10~T (well above the SO-hf
crossover) due to a very weak magnetic field dependence of the SO component for electrons. 

\section{Conclusions}
\label{sec:concl}

We have calculated the rates of phonon-assisted hyperfine and spin-orbit induced spin
flips during electron and hole 
relaxation in a self-assembled QDM using a~multi-band theory of hyperfine couplings based on the \kp
model. We have predicted a cross-over between the two processes as dominant 
spin-flip mechanisms at magnetic fields of a few Tesla and on the order of $0.1$~T for
electrons and holes, respectively, with the hf mechanism scaling as $1/B^{2}$ and
dominating at lower fields over the nearly magnetic-field-independent SO channel.
For the QDM structure considered here, the probability of spin flip during electron
tunneling between the QDs can be large at low 
fields (about 1\% for electrons and holes at magnetic fields of 0.1~T and 0.01~T,
respectively) but decreases strongly with increasing field, reaching values around $10^{-6}$
and $10^{-4}$ for electrons and holes, respectively, when the SO coupling dominates
the relaxation.

The interplay between the two channels with opposite magnetic-field dependence of the
relative spin-flip rates leads to a non-monotonic dependence of the total relative
spin-flip rate. The resulting minimum is particularly pronounced for holes. This prediction
relies on the assumed strong transverse hf coupling for holes, in line with some of the
recent experiments and with theoretical predictions based on the $d$-shell admixture to
the valence band states. As the spin-flip efficiency is in principle measurable in optical
experiments, this prediction might be used for testing the presence of the transverse hf
couplings. 

\acknowledgments
The authors acknowledge support by the
Polish National Science Centre under Grant No.~2014/13/B/ST3/04603 (PM, KG) and Grant
No.~2016/23/G/ST3/04324 (PK).
Calculations have been carried out using resources provided by
Wroclaw Centre for Networking and Supercomputing (\url{http://wcss.pl}), Grant No. 203.

%\bibliographystyle{prsty}
%\bibliography{library}

\end{document}